# *Operando* optical tracking of single-particle ion dynamics and phase transitions in battery electrodes

Alice J. Merryweather[1,2], Christoph Schnedermann[1,*], Quentin Jacquet[2], Clare P. Grey[2,*] & Akshay Rao[1,*]

[1] Cavendish Laboratory, University of Cambridge, J. J. Thomson Avenue, Cambridge CB3 0HE, United Kingdom
[2] Department of Chemistry, University of Cambridge, Lensfield Road, Cambridge CB2 1EW, United Kingdom

**Key to advancing lithium-ion battery technology, and in particular fast charging capabilities, is our ability to follow and understand the dynamic processes occurring in operating materials under realistic conditions, in real time, and on the nano- to meso-scale. Currently, *operando* imaging of lithium-ion dynamics requires sophisticated synchrotron X-ray or electron microscopy techniques, which do not lend themselves to high-throughput material screening. This limits rapid and rational materials improvements. Here we introduce a simple lab-based, optical interferometric scattering microscope to resolve nanoscopic lithium-ion dynamics in battery materials and apply it to follow the repeated cycling of the archetypal cathode material $Li_xCoO_2$. The method allows us to visualise directly the insulator-metal, solid solution and lithium ordering phase transitions in this material. We determine rates of lithium insertion and removal at the single-particle level and identify different mechanisms that occur on charge vs. discharge. Finally, we capture the dynamic formation of domain boundaries between different crystal orientations associated with the monoclinic lattice distortion at around $Li_{0.5}CoO_2$. The high throughput nature of our methodology allows many particles to be sampled across the entire electrode and, moving forward, will enable exploration of the role of dislocations, morphologies and cycling rate on battery degradation. The generality of our imaging concept means that it can be applied to study any battery electrode, and more broadly, systems where the transport of ions is associated with electronic or structural changes, including nanoionic films, ionic conducting polymers, photocatalytic materials and memristors.**

Lithium-ion batteries have emerged as the frontrunner technology to achieve high-power, intermediate-scale energy storage, with a broad range of applications including electric vehicles and portable devices. A major challenge associated with the development of improved batteries is to understand and optimise the processes by which lithium-ions intercalate into the active host materials. *Operando* characterisation techniques are needed to examine the fundamental limits to rate performance in a working battery environment.[1] However, tracking driven ionic motion in battery materials using established electrochemical methods is a highly challenging task, since the active particles are often intrinsically disordered at both the particle and electrode level, and can behave heterogeneously.[2,3] Advanced *operando* synchrotron-based[3–9] and electron microscopy measurements[10,11] can probe the length and time scales required to examine individual active particles, providing chemical and structural information. For example, *operando* transmission X-ray microscopy has revealed that non-uniform intercalation in $Li_xFePO_4$ (LFP) causes compositional heterogeneity within particles to be enhanced during delithiation and supressed during lithiation.[4] However, such techniques are costly and time-intensive, suffer from beam-induced sample degradation, and often require highly specialised cell geometries.[12]

Here, we establish optical interferometric scattering microscopy (iSCAT)[13–17] as a rapid and low-cost imaging platform to visualise and quantify ion dynamics at the single-particle level. We apply iSCAT to study $Li_xCoO_2$ (LCO, $0 < x < 1$), the archetypal layered cathode material which adopts the rhombohedral $\alpha$–$NaFeO_2$ structure (see Extended Data Figure 2a).[18,19] Despite being almost ubiquitous in portable electronics, the dynamics of phase transitions in LCO are not well understood at the single-particle level: for instance, the reported lithium-ion diffusion coefficients vary over six orders of magnitude,[20–26] and the degree of compositional heterogeneity within single particles is little explored.[7,26]

*Operando* iSCAT studies were carried out using an optically-accessible half-cell (Figure 1a). The working electrode was comprised of single-crystal LCO particles (~10 μm average size, Figure 1b), sparsely dispersed in a matrix of nanoparticulate carbon and polymer binder, pressed onto an aluminium mesh current collector.

The galvanostatic performance of the active electrode in the optical cell, at a rate of 2C, (Figure 1c,d) is in good agreement with previous reports[27] and with performance in a coin cell (Extended Data Figure 1). In the differential capacity (d$Q$/d$V$) plots, prominent peaks are resolved at ~4.0 V (delithiation) and ~3.8 V (lithiation) (I and IV, Figure 1d), which are associated with a biphasic (first-order) transition from the semiconducting lithium-rich phase (approximate composition, $Li_{0.95}CoO_2$) to the metallic delithated phase (approximate composition, Li$_{0.75}$CoO$_2$; see Supplementary section 5).[28,29] Smaller peaks in the differential capacity plots (II and III, Figure 1d) are associated with the lithium ordering transition at a composition corresponding to Li$_{0.5}$CoO$_2$ and the formation of a monoclinic phase.[27] The sloping regions between peaks I-II and III-IV correspond to a solid solution mechanism, during which the lithium content and cell parameters change continuously.

iSCAT images of a single LCO particle were obtained during five electrochemical cycles at 2C (Figure 1c,d). Full details of the iSCAT methodology can be found in the Methods section. Briefly, as illustrated in Figure 1e,f, following widefield illumination at 785 nm ($\mathbf{E}_i$), the objective collects back-scattered light from the particle ($\mathbf{E}_s$), as well as reflected light from the glass window/electrolyte interface ($\mathbf{E}_r$). Both components are subsequently imaged onto a camera, where they interfere to produce the observed iSCAT image (Supplementary section 1). The resultant iSCAT intensity is determined by the local dielectric properties of the sample material.[13–17] We take advantage of the fundamental correlation between the local lithium content in LCO and the local electronic structure (and thus the local dielectric properties),[30,31] which controls the scattering intensity. This allows us to probe intercalation dynamics at the single-particle level, in real time with sub-5 nm precision (Supplementary section 4).[33] Critically, our approach works in the absence of optical absorption[32] and can be readily applied to examine multiple particles within the same electrode.

The raw iSCAT image of this representative LCO particle (~9.6 × 6.8 μm) shows a brightly scattering particle on top of a characteristic speckle pattern originating from scattering contributions from the surrounding carbon matrix (Figure 1g). The relatively spatially-uniform intensity across the particle indicates a mostly-flat scattering surface, implying that the direction of observation is along the *c*-axis of this crystal structure, normal to the layers of CoO$_6$ octahedra (Extended Data Figure 2, Supplementary section 2), and ideally suited to investigate the in-plane ion transport within the layered host lattice.

As shown for one cycle in Figure 2a, the iSCAT intensity increased by 1.6 times during delithiation (0 - 25 min), followed by an equivalent decrease upon lithiation (25 - 50 min), indicating good reversibility. This confirms that the changes in dielectric properties caused by the underlying electronic structure changes in LCO (Supplementary section 1)[28] are sufficiently large to allow the delithiation and lithiation processes to be monitored.

Next, we investigate the spatially-resolved ion dynamics throughout the cycle (Figure 2b). At the beginning of the delithiation and during the sharp increase in cell potential to ~4.0 V, the particle intensity remained relatively constant and homogenous (A). However, from 3 - 12 min, (exemplified by B, 10.2 min) the iSCAT images showed a substantial degree of spatial inhomogeneity in the form of bright and dark features. Similar inhomogeneous features were seen from 39 - 48 min, (exemplified by G, 43.8 min), corresponding to an equivalent state of charge upon lithiation. Examination of the full iSCAT video (Supplementary Video 1) revealed that these bright and dark features propagated across the visible surface of the particle. The durations of these moving features are aligned with the biphasic transitions identified in the overall electrochemistry (red shaded, Figure 2a), allowing us to assign the moving features to propagating phase boundaries between the two stable phases (discussed below). Short-lived propagating features were also observed at ~23 min and equivalently at ~27 min (blue shaded, Figure 2a) which align excellently with the lithium ordering transitions and are further discussed below. Outside these transitions, *i.e.* during the solid solution mechanism, we observed a nearly spatially homogenous intensity change (Figure 2b, C-F).

A further 15 particles across multiple electrodes were examined. These studies yielded similar results, with an increase (decrease) in intensity upon delithiation (lithiation) and the observation of propagating phase boundaries, suggesting that the described behaviour is general across the electrode(s). Results for a second particle are presented later in the text, and included in Supplementary Videos 6-10.

Having established how iSCAT can reveal different mechanisms on the single particle level, we now turn to examine the biphasic insulator-metal transition in more detail (red shaded, Figure 2a), where the semiconducting $Li_{0.95}CoO_2$ phase and the metallic $Li_{0.75}CoO_2$ phase coexist. To visualise the phase boundaries, we extracted normalised sequential differential images, which represent the fractional intensity change over a 20 s duration (see Methods for detailed discussion).

During delithiation, intensity changes initially occurred at the particle edges (4.3 min, Figure 3a). After a lag time of ~3 min, new features emerged and spread across the bulk of the particle, originating predominantly from the bottom edge of this particular particle, and developing into a ring-like structure (10.0 min, Figure 3a). This ring feature progressively reduced in size and vanished at the end of the biphasic transition (11.0 – 11.7 min, Figure 3a). The ring is assigned to the observable phase boundary between the initial lithium-rich phase in the middle and the newly formed lithium-poor ($Li_{0.75}CoO_2$) phase growing inwards from the edges. The observed movement of the phase boundary is consistent with a so-called 'shrinking core' mechanism.[34,35] This behaviour was found in all cycles (Extended Data Figure 3). Our methodology allows us to extract the velocity of the propagating phase boundary, for which the average is ~20 nm s$^{-1}$ (at this 2C cycling rate), but as fast as 37 nm s$^{-1}$ at the end of the biphasic transition (Supplementary section 4).

Intriguingly, during lithiation we observed a different behaviour whereby a region of higher intensity first appeared in the top-right corner of the particle (39.3 min, Figure 3b) and then spread across the whole particle (40.0 – 44.0 min, Figure 3b). This process is best described as an 'intercalation wave' mechanism, where a single phase front (or small number of fronts) originating from one (or a small number of) nucleation point(s) moves across the particle.[34,36] While this mechanism occurred in all cycles, significant variations were found between cycles, both in the location of the first nucleation point and in the path taken by the new phase (Extended Data Figure 3).

Dynamic inhomogeneity between particles in an electrode can occur if the instantaneous current densities experienced by individual particles do not match the overall applied C-rate. For example, it has been shown that, during the biphasic process in LFP, only a small fraction of particles may be active at any given time,[9,37] leading to temporarily higher C-rates at the single-particle level. Therefore, to investigate the mechanism of the biphasic reaction in LCO further, the phase fractions in the particle were calculated and then used to derive the effective C-rate for the particle under observation (see Methods). Upon delithiation (Figure 3c), the biphasic reaction commenced at similar rate to that of the overall electrode (2C), with the new phase growing in from the particle edges and corners, to transform ~10% of the particle to approximately $Li_{0.75}CoO_2$ (4.3 min, Figure 3a). The single-particle C-rate then dropped, remaining low for ~3 min, before accelerating rapidly as the phase boundaries moved to form a shrinking ring, finally reaching a C-rate of almost 10C at the end of the biphasic transition. This delithiation behaviour was consistent across all cycles. Upon lithiation (Figure 3d), the new phase filled the particle at a C-rate oscillating around 2 – 4C for the selected cycle, but with significant variations between cycles which are associated with the phase boundaries travelling along different paths through the particle. (See Extended Data Figure 3 and Supplementary Videos 2-3 for sequential contrast images of the phase boundaries for all cycles).

A similar LCO particle from another electrode was monitored at applied C-rates from C/2 to 6C to explore how the mechanisms vary with rate (raw iSCAT and scanning electron microscopy images are reported in Extended Data Figure 2c). During delithiation at C/2, the integrated single-particle scattering intensity (Figure 4a) showed a peak during the biphasic transition (0-75 mA h g$^{-1}$) and a linear increase during the solid solution transition (75-155 mA h g$^{-1}$), with very similar behaviour during the following lithiation. At all applied C-rates, the intensity changes during the solid solution reaction remained linear (with time) and reversible. However, as the current density was increased, the peak in intensity associated with the biphasic transition shifted towards higher capacity during delithiation and lower capacity during lithiation, suggesting that the biphasic reaction in the observed particle lagged behind the ensemble electrochemistry. To understand this delay better, the single-particle C-rate during the biphasic reaction was extracted for each applied electrode current (Figure 4b,c). For both delithiation and lithiation, the single-particle C-rate increased sharply with the electrode current, reaching 25C during delithiation and 17C during lithiation at the highest current density (6C applied over the whole electrode). Figure 4d shows the progression of the new phase through the particle: the phase boundary progressed in an intercalation wave mechanism during lithiation at all applied C-rates, while a shrinking core mechanism occurred during delithiation at 2C, 4C and 6C, consistent with the first particle described above (Figure 3). At C/2 and 1C, delithiation appeared to follow a hybrid of the two mechanisms, with the new phase nucleating at two corners of the particle and propagating to finish at an edge - as opposed to the centre - of the particle (see Supplementary Videos 6-10).

The shrinking core mechanism – seen here on delithiation – is a consequence of the higher lithium flux across the active electrochemical surface (*i.e.* lithium-ion insertion/extraction, quantified via the charge transfer reaction rate) as compared to the lithium flux inside the particle (quantified via the lithium-ion diffusion rate).[35,36] This mechanism is therefore the result of a 'diffusion-limited' process.[35,36] In contrast, the intercalation wave mechanism is 'charge transfer-limited' and results in the formation of a phase front at equilibrium, which can have different morphologies depending the material/particle properties but features a reduced interfacial-area, which propagates across the particle as the reaction proceeds.[36] Our observation that the shrinking core mechanism observed during high-rate delithiation appears to switch towards an intercalation wave mechanism at lower rates is consistent with the decrease of the charge transfer rate at lower currents. This is in line with recent simulations on LCO, and experimental work on large graphite particles (>100 μm)[34,38]. However, this previous work makes no distinction between the mechanisms on delithiation and lithiation.

To explore the origins of the observed differences between lithiation and delithiation, we conducted phase field modelling[39] to identify parameters which control the phase boundary movement (methods detailed in Supplementary section 5). While being a highly simplified model, our results suggest that the difference between the delithiation and lithiation mechanisms can be explained by carefully considering the phase in which the charge transfer reaction occurs and the effect of the much lower lithium-ion diffusion coefficient in the lithium-rich phase ($Li_xCoO_2$, $1 \geq x > 0.95$) compared to the lithium-poor phase ($x < 0.75$):[20,40] During rapid delithiation, once the lithium-poor phase has nucleated, the charge transfer reaction proceeds at both the lithium-rich and lithium-poor surfaces of the particle. Due to the low diffusion coefficient in the lithium-rich phase, the delithiation in this phase is diffusion-limited and the lithium-poor phase builds up around all the active surfaces of the particle, resulting in a shrinking core mechanism (Supplementary Figure 6). During lithiation, although charge transfer occurs in both phases initially, the lithium concentration will rapidly build up to saturation in the lithium-rich phase and shut down this reaction pathway. Therefore, most of the charge transfer during lithiation will proceed via the high ionic mobility lithium-poor phase, leading to the charge transfer-limited intercalation wave mechanism for a large range of C-rates, as confirmed by our experiments (Supplementary Figure 7). Until now, there has not been a simple experimental method to observe these changes in the biphasic intercalation mechanism and connect them to parameters such as lithium diffusivity and single-particle C-rate.

The two observed LCO particles were (de)lithiated at maximum single-particle C-rates of 10C (for the particle in Figure 3c-d, 2C applied over the whole electrode) and 25C (for the particle in Figure 4b-c, 6C applied), requiring a high lithium diffusion coefficient in the material. To calculate a lower bound for the diffusion coefficient, delithiation of both particles during the biphasic process was simulated using various concentration-independent diffusion coefficients (Supplementary Figures 8-10). The resulting velocities of the phase boundaries were extracted and compared to the experimentally observed phase boundary velocities (~20 nm s$^{-1}$ at the 2C applied rate, Supplementary section 4). Good agreement was achieved using a chemical diffusion coefficient of the order of ~$10^{-9}$ cm² s$^{-1}$ or higher, in agreement with higher values estimated previously via theoretical[20] and muon spectroscopy[21] investigations (Supplementary Figure 11). Our single-particle results thus highlight the high-rate capabilities of LCO.

We now turn our attention to the lithium ordering transition (blue shaded, Figure 2a). Upon delithiation, lithium ordering causes a reduction in symmetry of the unit cell from rhombohedral to monoclinic, breaking the three-fold symmetry. Early *operando* XRD reports described this process as a second-order transition,[27] but more recent synchrotron studies show coexistence of the rhombohedral and monoclinic phases, suggesting a biphasic transition.[41]

To identify the overall changes in the particle during the ordering transitions, we computed the total normalised differential iSCAT images (*i.e.* the fractional intensity change for each pixel over the transition) for the first and fourth cycles (see Extended Data Figure 4 for all other cycles). For both cycles (Figure 5a,b), we observed intensity changes which were inversed between delithiation and lithiation, indicating a fully reversible ordering transition, as expected. In cycle 1 (Figure 5a), the particle intensity increased (decreased) relatively homogeneously upon ordering (disordering). On the other hand, in cycle 4 (Figure 5b), the ordering transition produced bright sharp lines in the particle with a three-fold symmetry, which disappeared again upon disordering.

We attribute the observed changes to the orientation of the newly formed monoclinic crystal lattice induced by the ordering transition. The absence of pronounced scattering lines in the bulk of the material in cycle 1 (Figure 5a) suggests that the monoclinic phase orientation was the consistent across the particle (Figure 5c), *i.e.* there is only one monoclinic domain present. By contrast, the appearance of bright lines with a three-fold symmetry in cycle 4 (Figure 5b) suggests the presence of three micron-sized ordered monoclinic domains, oriented at 120° with respect to each other, and which can be distinguished by brightly scattering domain boundaries (Figure 5d).

To follow the evolution of these structures in real time, we analyse the normalised sequential differential images (5 s frame interval) for the transition during lithiation (see Supplementary Videos 4-5, including delithiation). In cycle 1 (Figure 5e), two phase fronts are identified which emerge from opposite sides of the particle and approach each other head-on, travelling at a velocity of ~70 nm s$^{-1}$. By comparison, in cycle 4 (Figure 5f), the disordered phases grow in from three different locations separated by the visible domain boundaries. It should be noted that the phase boundaries move significantly faster than for the insulator-metal biphasic transition, largely because they involve long-range ordering of lithium-ions, rather than a (significant) change in lithium-ion concentration.[27]

Our work builds on the previous *ex-situ* electron diffraction observation of several ordered domains with three distinct orientations of the monoclinic phase within a single particle,[42] by observing the dynamics of the domain formation in real time. Here, we highlight that for a phase transition involving symmetry breaking and originating from multiple nucleation points, the new phases cannot readily fuse together if their orientations differ. The particle retains some memory of the nucleation conditions, in the form of domains. This is opposed to the case of the insulator-metal biphasic transition discussed above, for which symmetry is conserved across the phase transition, so that separate regions of the new phase can join together seamlessly. Considering the excellent cycling stability of LCO between 3.0 V and 4.2 V, the presence of the monoclinic domains at $Li_{0.5}CoO_2$ does not appear to be detrimental, probably because the monoclinic distortion leads to only a small deformation of the unit cell.[27] This would be quite different for the high voltage O3-O1 transition that occurs at a composition of approximately $Li_{0.15}CoO_2$, during which there are much larger changes to the cell parameters.[19]

**Conclusions**

In conclusion, we have established iSCAT microscopy as a powerful tool to track and quantify phase transitions in LCO on the nanoscale, in real time and under realistic operating conditions. The solid solution, biphasic and lithium ordering transitions were clearly resolved and correlated to the ensemble electrochemistry. Mechanistically, we were able to identify a preference for a shrinking core mechanism during delithiation and an intercalation wave mechanism during lithiation in the biphasic transition. These observations were rationalised in terms of the differences in lithium diffusivity in the two phases, with support from phase field modelling. The single-particle C-rates and phase boundary velocities were extracted, showing that individual particles are capable of sustaining much higher C-rates than those of the ensemble. The implication of this result is that an appropriately structured LCO electrode, where lithium transport in the electrolyte to the particle is not rate limiting, is capable of sustaining extremely high rates. Additionally, we observed the real-time formation and destruction of domains in the monoclinic structure around $Li_{0.5}CoO_2$.

We expect that the ability of this scattering microscopy methodology to provide real-time insights into nanoscale electronic or structural phase transitions, coupled with its straightforward lab-based implementation, will see it become an indispensable tool for high-throughput material discovery and mechanistic studies, complementary to existing synchrotron-based methodologies. For example, future work could examine the effects of grain boundaries and crystal defects[8] on phase transitions and ion intercalation mechanisms. Critically, the principle of using light-scattering to probe electronic structure changes is broadly applicable to a wide range of materials and promises to be generally valuable for the study of ferroelectrics, nanoionics, bioelectronics, photocatalytic materials, and memristors, as well as batteries.

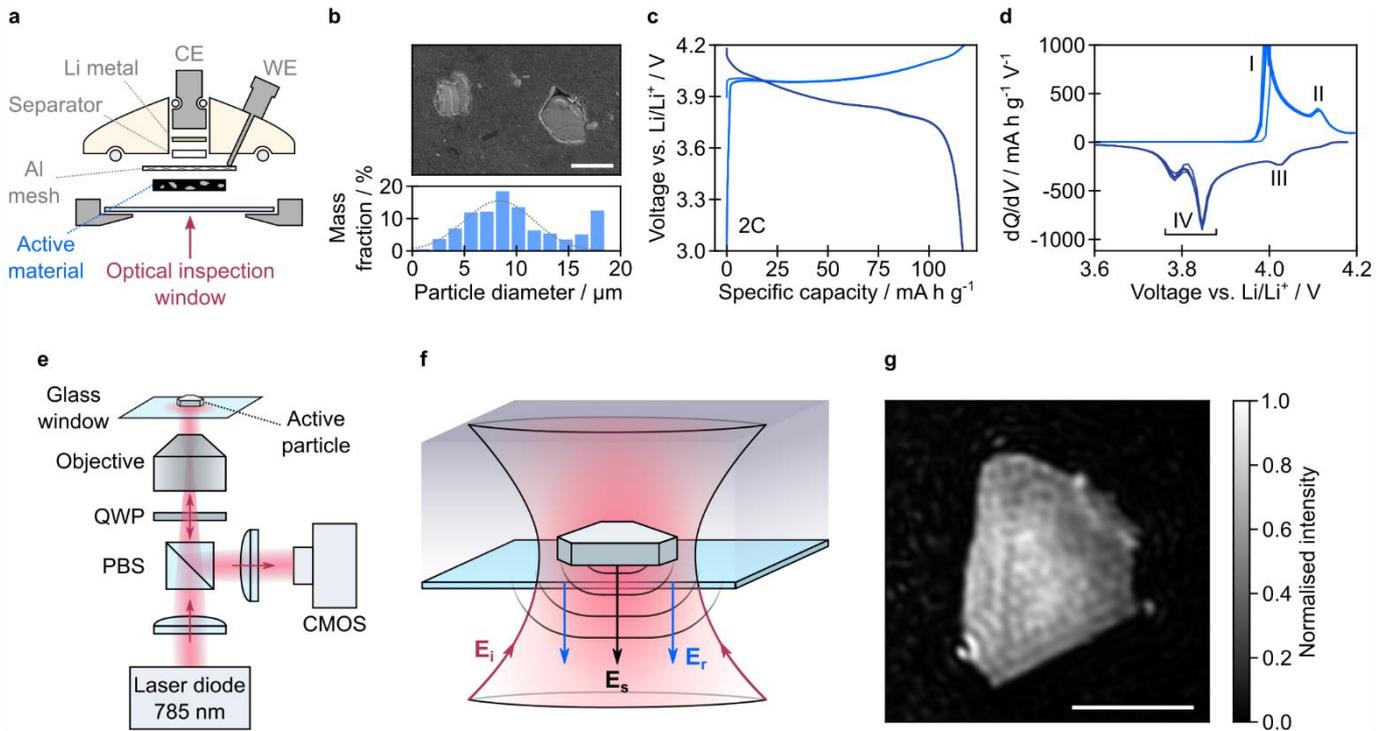

**Figure 1: Electrochemical performance and interferometric scattering microscopy of a LCO electrode.**
**a** Geometry of the optical microscopy half-cell. (WE = working electrode, CE = counter electrode). The counter electrode was lithium metal, the separator was glass fibre, and the cell stack was wet with standard carbonate liquid electrolyte (LP30). **b** Top: Scanning electron microscopy image of a dilute working electrode, showing two particles of LCO dispersed in a conductive matrix. Scale bar is 10 μm. Bottom: Mass-weighted diameter distribution for LCO particles (based on 681 particles). **c** Galvanostatic cycling (2C, 5 cycles) of the LCO electrode during *operando* optical measurements. **d** Corresponding differential capacity plots. The peaks attributed to biphasic transitions (I and IV) and lithium ordering (II and III) are indicated. **e** Optical setup of the interferometric scattering (iSCAT) microscope. (PBS = polarising beam splitter, QWP = quarter-wave plate, CMOS = complementary metal oxide semiconductor camera). **f** Schematic diagram of iSCAT signal generation. Incident light ($E_i$) is focussed onto an active particle of interest in the working electrode. The collected light includes a contribution scattered from the surface of the active particle ($E_s$) and a reference contribution reflected from the top interface of the glass window ($E_r$). **g** iSCAT image of a single active LCO particle in the electrode (250 μs exposure time). Intensity values are normalised to a range of 1. Scale bar is 5 μm.

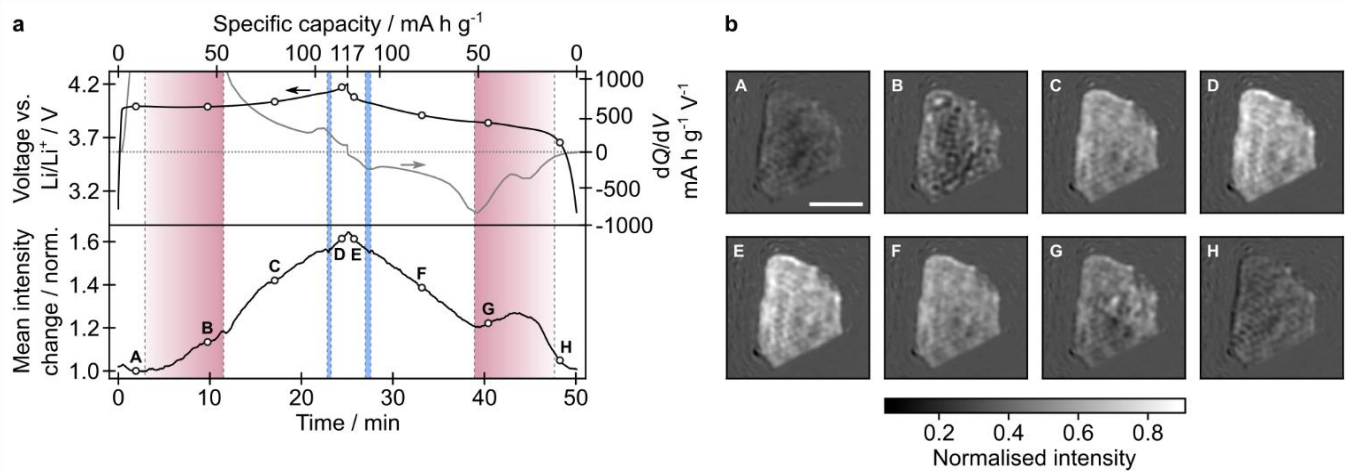

**Figure 2: Overview of the optical response of an active particle during battery operation.**
**a** Top: Galvanostatic and differential capacity plots shown in black and grey respectively, as a function of time (cycle 4, as plotted in Figure 1c and d). Bottom: iSCAT intensity change averaged over the active particle shown in Figure 1g, during galvanostatic cycling. Vertical red and blue shaded regions correspond to the durations of the biphasic and lithium ordering transitions, respectively, identified from images of this particle. **b** Background-subtracted iSCAT images of the active particle at the time-points indicated in panel a. Background subtraction was achieved by subtracting reference values for each pixel at the beginning of the cycle from the corresponding pixels in all subsequent images. Scale bar is 5 μm.

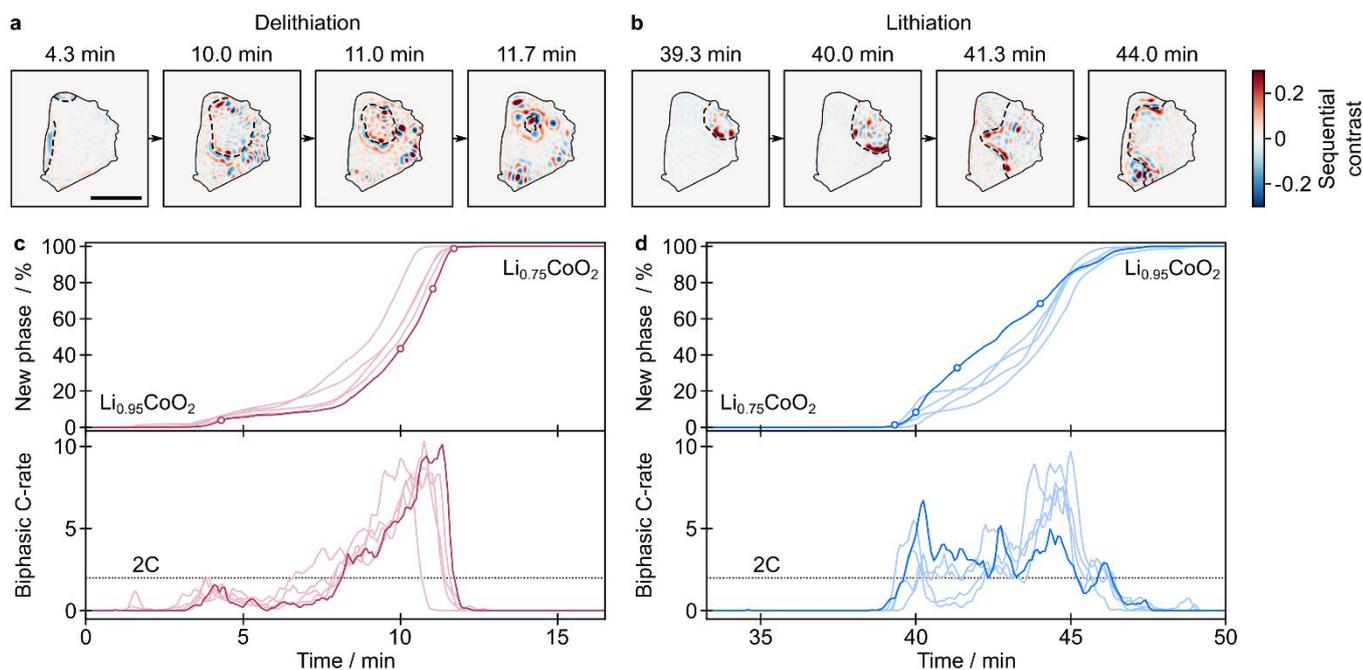

**Figure 3: Behaviour of biphasic phase transitions upon delithiation and lithiation.**
**a, b** Sequential differential images of the active particle upon delithiation and lithiation, respectively, during the biphasic transition (in cycle 4). The black dashed lines are a guide for the eye, representing the phase boundary position. Sequential contrast is obtained by dividing pixel intensity values by those from 20 s earlier, then subtracting 1, to represent the intensity changes over this timescale. Scale bar is 5 μm. **c, d** Top: Phase fraction of $Li_{0.75}CoO_2$ (delithiation) and $Li_{0.95}CoO_2$ (lithiation), respectively, as a function of time. All 5 cycles are shown, and time is measured from the start of each cycle. Time-points corresponding to the images in **a** and **b** are indicated as open circles on the traces for cycle 4 (darker shade). Bottom: The instantaneous single-particle C-rate for the biphasic transition, as obtained from the change in phase fraction.

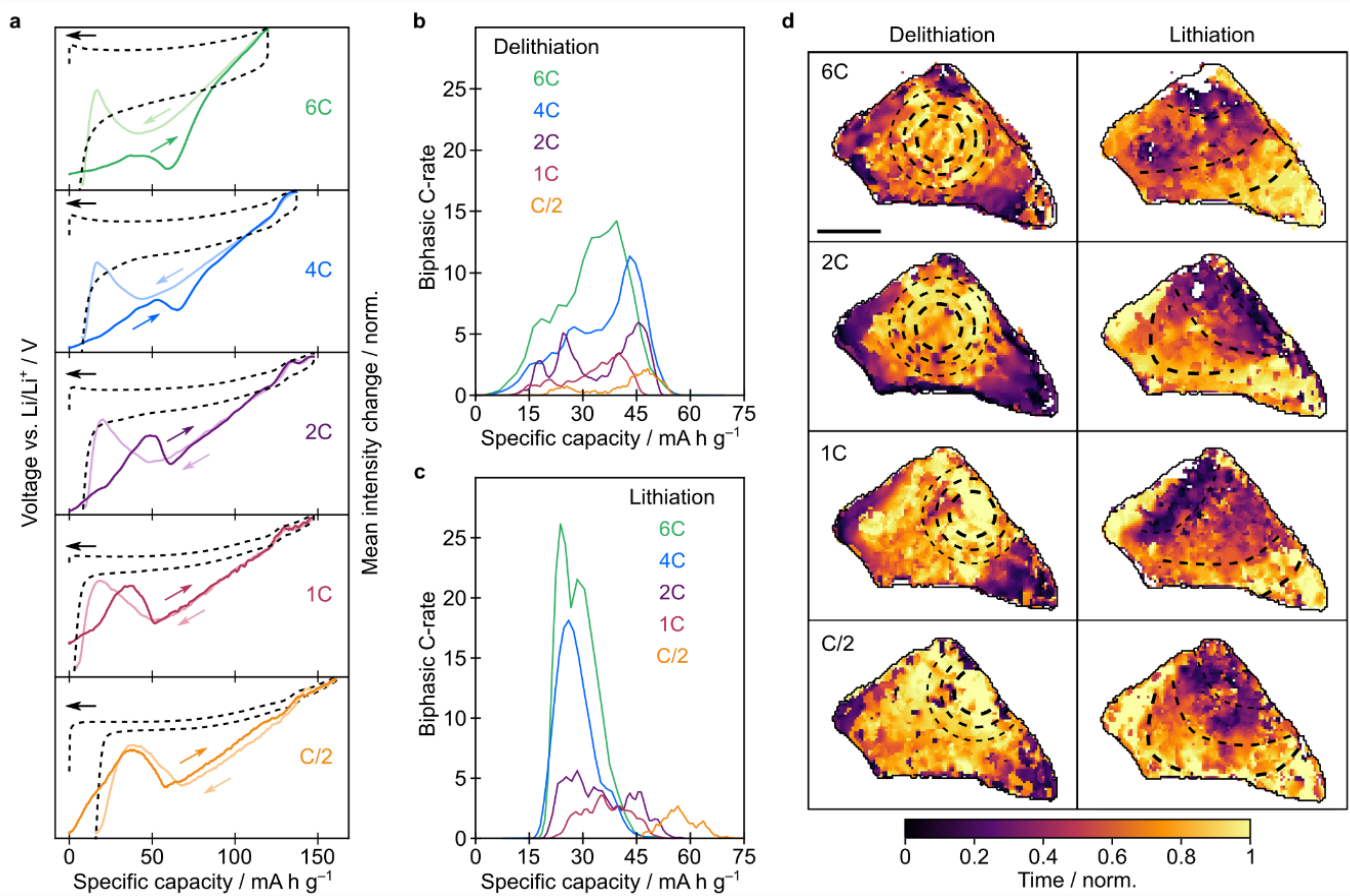

**Figure 4: Behaviour of biphasic phase transitions at various applied current densities.**
**a** Galvanostatic cell capacity plots with a voltage window of 3.0 V to 4.3 V (black dashed) for applied C-rates ranging from C/2 to 6C, with iSCAT intensity averaged over a single active particle during delithiation (dark colour) and lithiation (light colour). Note that the apparent loss of columbic efficiency at slower cycling rates is the result of parasitic reactions (likely occuring at the high-surface-area working electrode), while the LCO electrochemistry remains reversible. **b, c** Instantaneous single-particle C-rates for the biphasic transition during delithiation and lithiation, respectively. **d** Progression of the phase boundary through the active particle during the biphasic transition, for delithiation and lithiation. The colour scale represents the time at which each pixel experienced the phase boundary, as a fraction of the total duration of the biphasic transition. Solid black lines clarify the observed outline of the particle. Black dashed circles and lines are a guide to the eye to visualise the progression of the phase boundary. Scale bar is 2 μm.

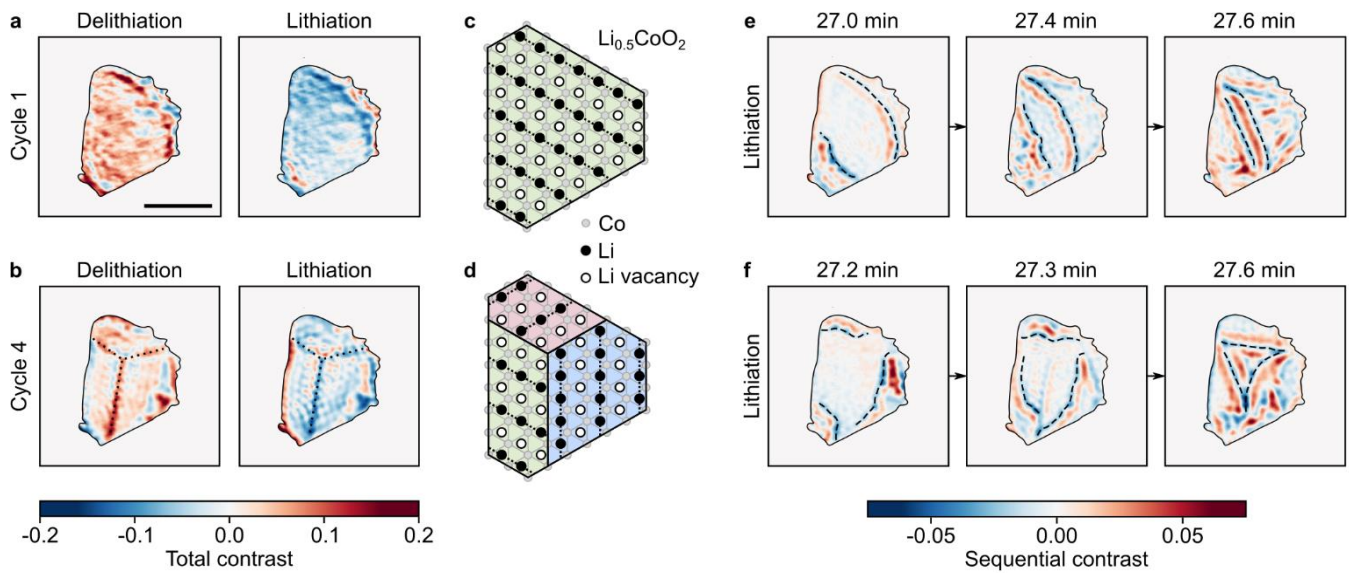

**Figure 5: Dynamics of the monoclinic distortion at Li$_{0.5}$CoO$_2$, with and without domain formation.**
**a, b** Images showing the total contrast resulting from lithium ordering in cycle 1 and cycle 4, respectively, for delithiation and lithiation. These were obtained by dividing pixel intensity values immediately after the transition by those from immediately before the transition, then subtracting 1, to represent the total intensity change caused by the transition. Scale bar is 5 μm. For cycle 4, the formation of the ordered state produces domain-like features, with three regions separated by bright lines at approximately 120°. **c, d** Schematics of a particle of Li$_{0.5}$CoO$_2$ with lithium-ions ordered into rows. The cases for a single monoclinic domain and for three monoclinic domains (with orientations of the rows differing by 120°) are shown, respectively. **e, f** Differential images for cycle 1 and cycle 4, respectively, during the transition causing lithium disordering upon lithiation. Sequential contrast is obtained by dividing pixel intensity values by those from 5 s earlier, then subtracting 1, to represent the intensity changes over this timescale.